# WORLDLINE DESCRIPTION OF THE ISGUR-WISE FUNCTION *

N. G. STEFANIS †

*Institut für Theoretische Physik II, Ruhr-Universität Bochum*
*D-44780 Bochum, Germany*

A. I. KARANIKAS ‡ and C. N. KTORIDES §

*University of Athens, Department of Physics, NPPS, Panepistimiopolis*
*GR-15771 Athens, Greece*

An effective field theoretic description of the Isgur-Wise function in heavy-meson transitions is presented which emulates soft interactions by a fermion worldline with an infinitesimal self-intersecting loop. A point-splitting regularization technique is used, which replaces pointlike worldlines by "ribbons" in the sense of Witten. The calculated vertex function is correctly normalized, does not depend on the heavy-quark mass, and complies with different sets of recent experimental data of the ARGUS and CLEO collaborations.

---

## 1   Introduction

In recent years a whole body of knowledge has been developed about the weak decays of hadrons containing a single heavy quark (see, e.g., Ref. 1). The strong interactions of such a heavy quark with light quarks and gluons can be described by an effective field theory which is invariant under changes of the flavor and the spin of the heavy quark and involves a single universal vertex function (form factor), introduced by Isgur and Wise, [2] which contains the strong interactions of quarks and gluons. This function is independent of the heavy quark mass and depends only on (the invariant product of) the four-velocities of the initial and final state mesons. Lacking an analytical understanding of the confinement dynamics, the Isgur-Wise function cannot be calculated from first principles within Quantum Chromodynamics (QCD). Current nonperturbative calculations utilize QCD sum rules or lattice simulations, employing heavy-quark symmetry.

In this paper we report on an alternative computation [3] which attempts at emulating the Isgur-Wise form factor by a universal vertex function, responsible for the elastic scattering of an infraparticle. [4] Technically, this is achieved by recasting the field theoretical system into particle-based language, [5,6] and employing in the evaluation of the particle path integral a self-intersecting worldline (contour) with an infinitesimal loop. The latter serves to simulate the transport of the intact soft boson "cloud" through the interaction point, and gives rise to an *exclusive* form factor. An introduction to the basics of this approach has been given in Ref. 7. Full accounts of the formalism were published in Ref. 8; more recent developments are discussed in Ref. 9.

## 2   Worldline Approach to the Isgur-Wise Function

A way as to make possible a first-principles calculation of the Isgur-Wise function relies in the present report on the assumption [9] that *global* characteristics of the soft QCD dynamics may be treated within an Abelian setup of infraparticle worldlines. *Local* aspects that are directly connected to the specific features of $SU(3)_c$ are left out, limiting the validity range of the method. On the other hand, this approach avoids typical problems encountered by *explicitly* resumming Feynman graphs and obtains results by *implicitly* taking into account an infinite number of such contributions within a low-energy *effective* theory. [8,9]

Let us now turn to the following generic process: an infraparticle propagates from point $x$ along a smooth worldline, say a straight line, with four-velocity $u_1(\tau)$ to point $z$, where it is suddenly derailed by a dynamical "pinch", and then proceeds to propagate on another smooth (straight) worldline, characterized by a four-velocity $u_2(\tau)$, until it reaches point $y$. Such a process can be adequately described by an appropriate three-point function $\Gamma(x, y, z)$ which accounts for the four-velocity change at $z$, and which labels initial and final states by four-velocities:

$$\Gamma(x,y,z) \stackrel{c \to 0}{=} \int_c^\infty dT \int_0^T ds \int [dp(\tau)] \, \text{tr} \mathcal{P} \exp\left\{ \int_0^T d\tau \, [ip(\tau) \cdot \gamma + m] \right\} G(x,y,z) \,, \quad (1)$$

where

$$G(x,y,z) = \int_{\substack{x(0)=x \\ x(s)=z \\ x(T)=y}} [dx(\tau)] \exp\left[ i \int_0^T d\tau \, p(\tau) \cdot \dot{x}(\tau) \right] \left\langle \exp\left\{ ig \int_0^T d\tau \dot{x}(\tau) \cdot A[x(\tau)] \right\} \right\rangle_A . \quad (2)$$

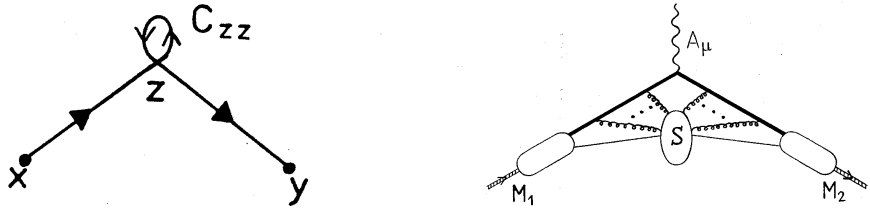

Figure 1: The left panel shows an infraparticle contour with an infinitesimal loop which serves to emulate the soft interactions during a heavy-meson transition. The latter is illustrated by the graph in the right panel.

Dissecting the particle's worldline into three disjoint branches $L_{xz^-}$, $C_{zz}$, and $L_{z^+y}$ (where $z^\pm = z \pm 0^+$ to denote that the crossing point $z$ has been excised), the expectation value of the Wilson line acquires the form $\left\langle \exp\left[ig \int_{L_{x,y}^z} dw_\mu\, A_\mu(w)\right]\right\rangle_A \left\langle \exp\left\{ig \int_{s_1}^{s_2} d\tau\, \dot{x}(\tau) \cdot A[x(\tau)]\right\}\right\rangle_A$. The contribution associated with the overlap of the soft boson "clouds" during the four-velocity transition process is encoded in the factor involving the loop $C_{zz}$ (left panel of Fig. 1), the latter to be entered and exited *smoothly* with four-velocities $u_1$ and $u_2$, respectively. Accordingly, the three-point function becomes the product of a purely kinematical factor $G(x, y|z)$ which describes coexisting in and out four-velocity states, described by infrapropagators, [13] and a factor $G^{(1)}(z, z)$ which is associated with a single loop that shrinks to zero (see for details Ref. 8). The analogous situation for the heavy meson transition in QCD is shown in the right panel of Fig. 1. Here the rearrangement of the light degrees of freedom during a four-velocity change gives rise to form-factor suppression, described by the *universal* Isgur-Wise function.

A uniform parametrization of the loop integral $\oint_{C_{zz}^{(1)}} dx_\mu A_\mu[x(\tau)]$ is provided by $\tau \to t = \tau - (s_1 + s_2)/2$ accompanied by the coordinate readjustment $x(\tau) \to x^c(t) = x\left[+(s_1 + s_2)/2\right]$, where $t$ is restricted in the interval $\left[-\frac{\tilde{\tau}}{2}, \frac{\tilde{\tau}}{2}\right]$ and $\tilde{\tau} = s_2 - s_1$. Then, all gauge-dependent terms vanish identically, as expected, and employing Witten's ribbon regularization, [10] the final result for the boson line exponential is [8]

$$I(\theta) \simeq - \left(g^2/4\pi^2\right) \ln\left(\tilde{\tau}/2b\right) \pi\theta/\sqrt{1-\theta^2}\,, \qquad (3)$$

where $\theta = u_1 \cdot u_2$. In order to disentangle IR from UV effects, we make a RG scale readjustment to tune the coupling constant to values pertaining to the IR domain: $g^2(b^2) \ln(\tilde{\tau}/2b) = g^2(\mu^2) \ln(\tilde{\tau}\mu) = g^2(\mu^2) \ln k$. The parameter $k$ may be viewed as relating long- and short-distance regimes. Performing renormalization, the result for the three-point function reads

$$\Gamma_{ren}^{(1)}(x, y, z) = -\exp\left(-\frac{g^2}{4\pi} \frac{\theta}{\sqrt{1-\theta^2}} \ln k\right) \Gamma_{ren}^{(0)}(x, y, z)\,, \qquad (4)$$

from which the nonperturbative part of the vertex function is obtained by removing kinematical contributions according to

$$F(x, y, z) = \frac{\Gamma_{ren}^{(0)}(x, y, z) + \Gamma_{ren}^{(1)}(x, y, z)}{\Gamma_{ren}^{(0)}(x, y, z)}\,. \qquad (5)$$

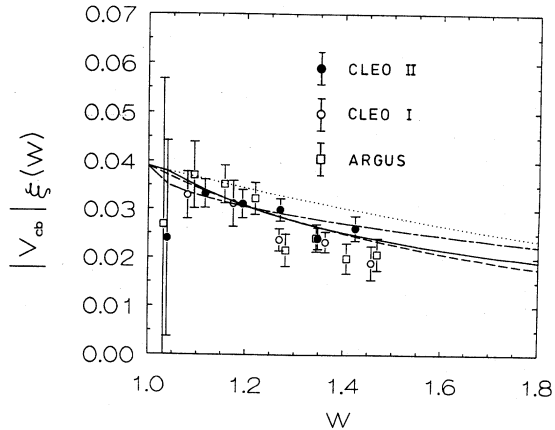

Figure 2: Selected theoretical predictions in comparison with recent experimental data of different collaborations indicated in the figure. The curves correspond to our result (solid line), and to those obtained by Isgur [11] (dotted line), Radyushkin [12] (dashed-dotted line), and Neubert [1] (dashed line).

The final result is the following *universal* (i.e., mass- and contour-independent) expression

$$F(x, y, z) = F(\theta) = 1 - \exp\left(-\frac{g^2}{4\pi} \frac{\theta}{\sqrt{1-\theta^2}} \ln k\right) , \qquad (6)$$

which can be analytically continued to Minkowski space [8] by replacing $\theta \to w \equiv \tilde{u}^1_\mu \tilde{u}^{2\mu} = \frac{1}{\theta}$, where $\tilde{u}^1$ and $\tilde{u}^2$ are the Minkowski counterparts of $u^1$ and $u^2$, respectively. The connection to the leading-order Isgur-Wise form factor is established by replacing $e^2$ by $(4/3)g_s^2$ and adapting the parameter $k$ to scales involved in heavy-meson decays by setting $k = m_Q/\bar{\Lambda}$, where $m_Q$ is the heavy-quark mass, and $\bar{\Lambda}$ characterizes the energy scale of the light degrees of freedom. In this way we arrive at

$$\xi(w) = 1 - \exp\left(-\frac{4}{3} \alpha_s \frac{1}{\sqrt{w^2-1}} \ln k\right) . \qquad (7)$$

The result of this first-principles computation is correctly normalized ($\xi(w=1) = 1$), and yields for the rather typical values $k \approx 8$ and $\alpha_s \simeq 0.21$ the model form

$$\xi(w) = 1 - \exp\left(-\frac{0.582}{\sqrt{w^2-1}}\right) . \qquad (8)$$

This vertex function is shown in Fig. 2 in comparison with other theoretical calculations together with the experimental values reported by different collaborations (see Ref. 8 for details and references). In the kinematic region accessible to semileptonic decays ($1 \leq w \leq 1.6$), Eq. (8) has physical characteristics close to those expected for the "true" Isgur-Wise function, except for the slope at zero recoil which turns out to violate the Bjorken bound. This is due to the Abelian setup, which is too simplifying a scheme to accomodate all nonperturbative phenomena of QCD.

## 3  Conclusions

The investigations presented here indicate that our low-energy effective approach, which is based on the quantum mechanical particle path integral rather than Feynman graphs in field theory, is useful for the nonperturbative calculation of quantities relating to soft interactions. The extension of the method to the ultraviolet regime, sketched in Ref. 9 in connection with fractal contours, seems also feasible. Work in this direction is in progress.

## References

1. M. Neubert, *Phys. Reports* **245**, 259 (1994).
2. N. Isgur and M. Wise, *Phys. Lett.* **B232**, 113 (1989); **B237**, 527 (1990).
3. A.I. Karanikas, C.N. Ktorides, and N.G. Stefanis, *Phys. Lett.* **B301**, 397 (1993).
4. B. Schroer, *Fortschr. Phys.* **11**, 1 (1963).
5. A. Kernemann and N.G. Stefanis, *Phys. Rev.* **D40**, 2103 (1989).
6. A.I. Karanikas and C.N. Ktorides, *Phys. Lett.* **B275**, 403 (1992); *Int. J. Mod. Phys.* **A7**, 5563 (1992); *Phys. Rev.* **D52**, 5883 (1995).
7. N.G. Stefanis, A.I. Karanikas, and C.N. Ktorides, in *Proceedings of the International Conference on Hadron Structure '94*, Košice, Slovakia, 1994, edited by J. Urbán and J. Vliáková (Košice, Slovakia, 1994), p. 134.
8. A.I. Karanikas, C.N. Ktorides, and N.G. Stefanis, *Phys. Rev.* **D52**, 5898 (1995).
9. N.G. Stefanis, in *International Conference on Problems of Quantum Field Theory*, Alushta, Crimea, Ukraine, May 13-18, 1996. Bochum preprint RUB-TPII-08/96 (June 1996), [hep-th/9607063, 8 July, 1996].
10. E. Witten, *Commun. Math. Phys.* **92**, 451 (1984).
11. N. Isgur, *Phys. Rev.* **D43**, 810 (1991).
12. A.V. Radyushkin, *Phys. Lett.* **B271**, 218 (1991).
13. A.I. Karanikas, C.N. Ktorides, and N.G. Stefanis, *Phys. Lett.* **B289**, 176 (1992).